\documentclass[structabstract, rnote]{aa} 

\usepackage{color}
\usepackage{natbib}			
\usepackage{graphicx} 			
\usepackage[varg]{txfonts}		
\usepackage[english]{babel}		
\usepackage{nameref}			
\usepackage{paralist}			
\usepackage{multirow}

\begin{document}
  \title{Identifying the best iron-peak and $\alpha$-capture elements  \\  for chemical tagging: The impact of the number of lines on measured scatter}
  \titlerunning{Identifying the best elements for chemical tagging}

  \author{V. Adibekyan\inst{1}
	  \and P. Figueira\inst{1}
	  \and N. C. Santos\inst{1,2}
	  \and S. G. Sousa\inst{1}
	  \and J. P. Faria\inst{1,2}
	  \and E. Delgado-Mena\inst{1}
	  \and M. Oshagh\inst{1}
	  \and \\ M. Tsantaki\inst{1}
	  \and A. A. Hakobyan\inst{3}
	  \and J.~I.~Gonz\'{a}lez Hern\'{a}ndez\inst{4,5}
	  \and L. Su\'{a}rez-Andr\'{e}s\inst{4,5}
	  \and G. Israelian\inst{4,5}
	 }

  \institute{
  	  Instituto de Astrof\'isica e Ci\^encias do Espa\c{c}o, Universidade do Porto, CAUP, Rua das Estrelas, 4150-762 Porto, Portugal
  	  \and
	  Departamento de F\'isica e Astronomia, Faculdade de Ci\^encias, Universidade do Porto, Rua do Campo Alegre, 4169-007 Porto, Portugal
	  \and
	  Byurakan Astrophysical Observatory, 0213 Byurakan, Aragatsotn province, Armenia
	  \and
	  Instituto de Astrof\'{\i}sica de Canarias, 38200 La Laguna, Tenerife, Spain
	  \and 
	  Departamento de Astrof{\'\i}sica, Universidad de La Laguna, 38206 La Laguna, Tenerife, Spain
}

  \date{Received date / Accepted date }
 
  \abstract
  {}
  {The main goal of this work is to explore which elements carry the most information about the birth origin of stars and as such that are best suited for
  chemical tagging. }
  {We explored different techniques to minimize the effect of outlier value lines in the abundances by using Ni abundances derived for 1111 FGK type stars. We 
  evaluated how the limited number of spectral lines can affect the final chemical abundance. Then we were able to make an efficient  even footing comparison 
  of the [X/Fe] scatter between the elements that have different number of observable spectral lines in the studied spectra. }
  {We found that the most efficient way of calculating the average abundance of elements when several spectral lines are available is to use a weighted mean (WM) where as a
  weight we considered the distance from the median abundance. This method  can be effectively used without removing suspected outlier lines. 
  We showed that when the same number of lines is used to determine chemical abundances, 
  the [X/Fe] star-to-star scatter for iron group and $\alpha$-capture elements is almost the same. On top of this, but at a lower level the largest scatter was observed 
  for Al and the smallest for Cr and Ni.}
  {We recommend caution when comparing [X/Fe] scatters among elements that have a different number of spectral lines available. 
  A meaningful comparison is necessary  to identify elements that show the largest intrinsic scatter and can  be thus used for chemical tagging.}

  \maketitle
%
\section{Introduction}					\label{sec:Introduction}

Studies of large samples of stars are very important for understanding the Galactic and stellar chemical evolution. 
Understanding the effects of these two mechanisms is, in turn, crucial for the studies of chemical properties of individual stars.
A representative example is the so-called T$_{c}$-trend: a trend of chemical abundance with the condensation temperature of the elements, 
whose real nature is still under debate \citep[e.g.][]{Melendez-09, Ramirez-09, Jonay-10, Jonay-13, Schuler-11, Adibekyan-14, 
Onehag-14, Maldonado-15, Nissen-15}.

Precise and detailed chemical composition studies of large samples of stars are also of great importance for different venues of Galactic astronomy.
One of these venues goes towards a so-called chemical tagging technique: identifying stars with identical chemical properties. This technique was 
introduced by \citet[][]{Freeman-02}, and then explored and developed by many  other authors \citep[e.g.][]{DeSilva-06, Tabernero-12, Tabernero-14, Mitschang-13, 
Mitschang-14, Blanco-Cuaresma-15}. Chemical tagging is a very powerful tool to identify stellar groups and clusters \citep[e.g.][]{Tabernero-14, DeSilva-13, 
Spina-14a, Spina-14b, Quillen-15} and even to identify solar siblings \citep[e.g.][]{Batista-14, Ramirez-14, Liu-15}.

In all likelihood, not all elements are equally useful for chemical tagging. A way of selecting the elements that can be used to tag stars 
is to look at the star-to-star [X/Fe] abundance ratio scatter at solar metallicities, where the Galactic chemical evolution does not have a very strong
effect. Elements that show largest star-to-star scatter are the more informative, being of physical origin.

The works of \citet{DeSilva-06, DeSilva-07, DeSilva-09} on open clusters and those of \citet{Ramirez-09} and \citet{Jonay-10} for solar-twins/analogs
clearly show that most of the elements show very small star-to-star [X/Fe] scatter. These authors performed a fully differential 
chemical abundance analysis in a line-by-line basis with respect to a solar spectrum reference, as well as to a star which is expected to belong to a given open 
cluster or kinematical group. In particular, in the recent work of \citet{Ramirez-14}, a higher weight/priority 
 to Na, Al, V, Y, and Ba were given for chemical tagging. However, in all the mentioned studies when the star-to-star [X/Fe] scatters were compared for different elements, 
 an important parameter was not 
taken into account: the number of spectral lines used to derive abundances for each element. 

In this work, using a large and high-quality data of solar-type stars, we study the dependence of [X/Fe] scatter on the number of spectral lines. 
This allows us to make a  comparison on the same ground of the [X/Fe] scatter for different elements by
using the same number of lines. Our sample comes from \citet[][]{Adibekyan-12} and consists of 1111 FGK-type dwarfs observed with the high-resolution HARPS spectrograph.
The stellar parameters and abundances of the stars were derived from the high signal-to-noise ratio (SNR) spectra with a median SNR of 235 (only 15\% of the spectra
have SNR $<$ 100).

We organize our paper as follows.In Appendix\,\ref{app:a}, we  quantify the precision in the abundance value as a function of number of spectral lines and 
summarize the results in Sect.\,\ref{sec:abundances}. The discussion on the [X/Fe] star-to-star abundance scatter and 
conclusions are presented in Sects\,\ref{sec:scatter} and \ref{sec:conclusion}.

\section{Reducing the impact of outlier lines}	\label{sec:abundances}

The data used in this work was taken from \citet[][]{Adibekyan-12}, which provides chemical abundances for 12 iron-peak and $\alpha$-capture elements (15 
ionized or neutral species).
In the present paper we did not use the final (average) abundances of different elements, but instead we used the abundances derived from individual lines of each element. 
As stated previously, this is because we aim at studying the dependence of precision in abundances on the number of lines.

A standard, and widely used technique to calculated chemical abundances derived from several spectral lines is to apply an outlier removal criteria and then
calculate the arithmetic mean (AM) of the abundances from the remaining lines. However, the detection of outliers is not an easy task. There are several 
outlier removal methods discussed in the literature (e.g. $\sigma$-clipping \citep[e.g.][]{Shiffler-88}, 
modified Z-score \citep[][]{Iglewicz-93}, Tukey's (boxplot) method \citep{Tukey-77}, and median-rule \citep[][]{Carling-98}), however most of them are model dependent while 
depending on the applied threshold for which there is no clear prescription or theoretical ground. It is appropriate to note, that outlier removal is 
not the only method used to 
characterize an underlying distribution in a dataset. An example is the weighted least-squares regression to minimize the effects of outlier data \citep[][]{Rousseeuw-87}. 

In  Appendix\,\ref{app:a}, we present an comprehensive  discussion about different outlier methods and a new WM method where as weight we 
use (inverse) distance from the median value as measured in units of standard deviations (SD). We made several tests to evaluate the impact of 
outliers on the chemical abundances (using Ni for our tests), and the dependence of the precision of the abundances on the number of lines.

Our tests showed that when the number of lines is large, different outlier removal techniques and criteria provide similar final abundances. However, the 
line-to-line dispersion, which is usually used to estimate the error on abundances, strongly depends on the criteria and can be artificially 
(unrealistically) reduced depending on the outlier removal method and threshold. We conclude and recommend to use the WM 
 (instead of any outlier removal technique) when several lines are available at hand.

We found that even for solar-type stars for which high-quality data is available, significant deviations in abundances from the real 
value are possible when the number of lines is small.

We refer the interested reader to Appendix\,\ref{app:a}  for the details of the tests and discussion.

\begin{figure}
\begin{center}
\begin{tabular}{c}
\includegraphics[angle=0,width=1.0\linewidth]{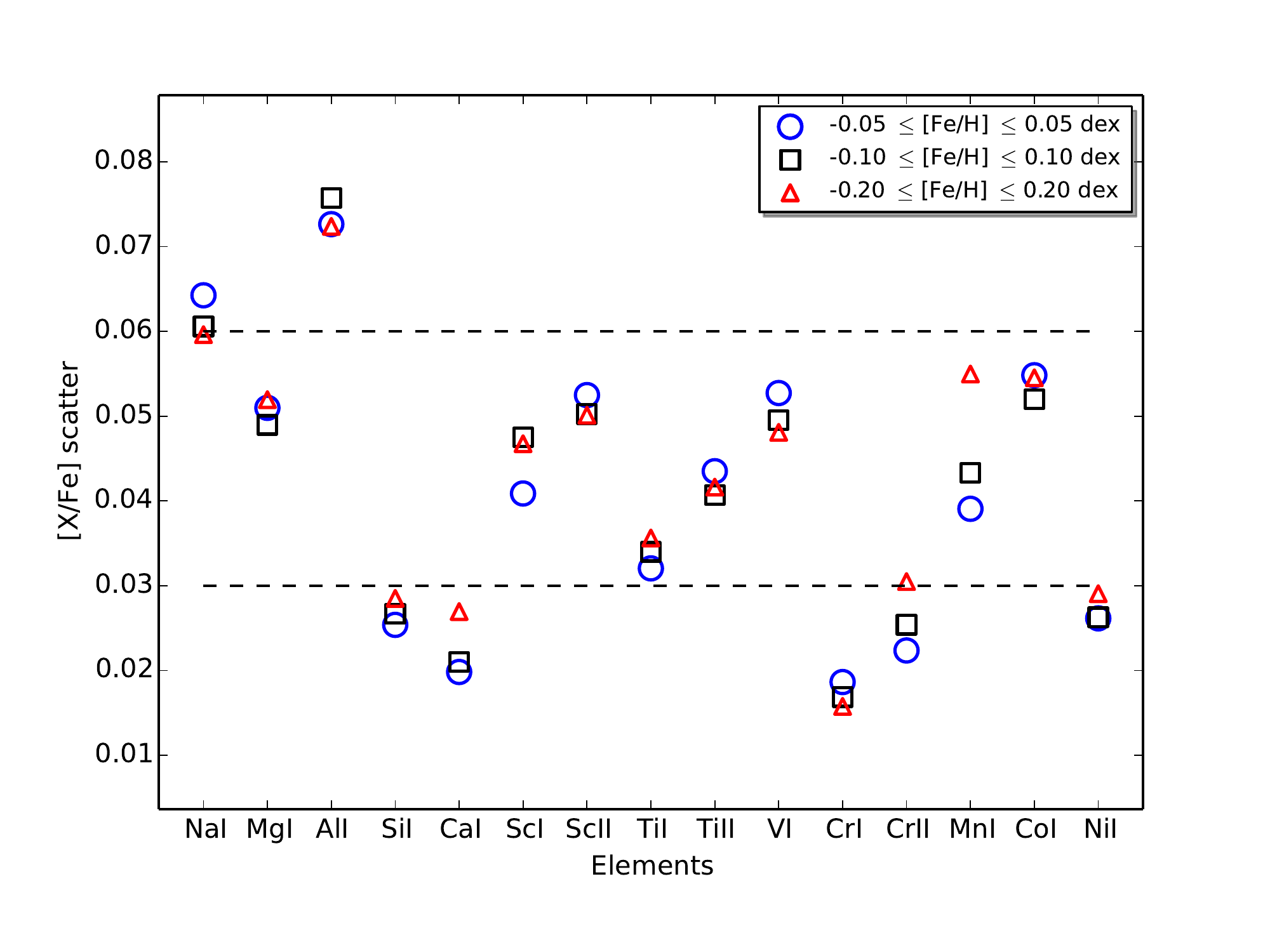}
\end{tabular}
\end{center}
\vspace{-0.5cm}
\caption{[X/Fe] star-to-star scatter for solar-analogs with solar-metallicity. The dashed lines, which represent [X/Fe] = 0.03 and 0.06 dex, 
are just to make the comparison of the [X/Fe] scatters between the elements visually easier.}
\label{fig_elfe_scatter_300k}
\end{figure}

\section{[X/Fe] star-to-star scatter}	\label{sec:scatter}

Recently, several works  on solar analogs \citep[e.g.][]{Ramirez-09, Jonay-10, Jonay-13}, showed that the [X/Fe] versus [Fe/H] trends show very
small star-to-star scatter at the solar metallicities for most of the elements. In Fig.~\ref{fig_elfe_scatter_300k}, we plot [X/Fe] scatter ($rms$) for dwarf 
stars ($\log g \geq$ 4 dex) that have effective temperatures within 300 K of that of the Sun (T$_{eff} \leq 5777\pm300$ K) and have metallicities 
in the range of [Fe/H] = 0.0$\pm$0.05, [Fe/H] = 0.0$\pm$0.10, and [Fe/H] = 0.0$\pm$0.20 dex, respectively. 
The abundances of all the elements were derived using all the available lines by applying the WM technique.
We selected only stars with solar metallicities to
minimize the effect of Galactic chemical evolution and the thin/thick disk dichotomy \citep[however see the discussion in][about thin/thick disk separation at
solar and super-solar metallicities]{Adibekyan-11, Adibekyan-13}.  The constrain on T$_{eff}$ serves to select the stars with the highest
precision of the stellar parameters and chemical abundances \citep[][]{Sousa-08, Adibekyan-12, Tsantaki-13}.
We note that the sample size is large enough to minimize the  errors related to the sampling of the population. For example,
if the scatter (standard deviation) is of about 0.05 dex (which is the case for most of the elements), the 95\% confidence interval of this value would be
from 0.045 to 0.056 for the sample size of 152 (the number of stars in the metallicity range of 0.0$\pm$0.10 dex)\footnote{
The confidence interval of SD can be calculated as presented in \citet[][]{Sheskin-07}.}.

Fig.~\ref{fig_elfe_scatter_300k} shows that the highest scatter is observed for Na and Al, and the [X/Fe] scatter for Si, Ca, Cr, and Ni is the lowest. 
The number of available lines that were used to derive abundances of Na and Al is the lowest: only two lines, while elements showing the smallest scatter 
usually have more than 10 lines. From the figure, it is apparent that the scatter does not change much when different metallicity intervals are used. The only 
exception is Mn where scatter increases with the width of the metallicity interval. We note, that for the derivation of Mn abundances
we did not consider hyperfine structure (hfs), which is important for odd-Z elements and, if not considered might overestimate the Mn abundances deduced from a given EW. 
This can be one of the reasons for the observed increase of [Mn/Fe] scatter with metallicity. Another reason for the observed high scatter at larger range of [Fe/H]
could be the strong Galactic evolution trend in the [Mn/Fe] -- [Fe/H] plane at solar metallicities \citep[e.g.][]{Adibekyan-12, Battistini-15}. We note,
that the trend is strong even if the hfs effect, but not a non-LTE effect is taken into account \citep[e.g.][]{ Battistini-15}.

To evaluate the impact of the number of lines (e.g. precision) on the [X/Fe] scatter we did a test similar to that presented in the previous section
(see Appendix\,\ref{app:a} for the details). For each element, 
we randomly drew $N$ number of lines (where $N$ is from one to the maximum number of lines) and calculated the [X/Fe] scatter for solar analogs in the 
metallicity range of 0.0$\pm$0.10 dex. If the number of possible combinations of the lines is less than 1000 we considered all of them, else we 
limited ourselves to fixed number of 1000 random (but different i.e., without replacement) combinations.

In Fig.~\ref{fig_elements_elfe_scatter_nlines}, we plot the dependence of [X/Fe] star-to-star scatter as a function of the number of lines that were used for [X/H] abundance 
derivations. The plot clearly shows that the average scatter decreases with the number of lines. The plot also shows that the WM  always
gives smaller scatter than the AM (of course, when the number of lines is larger than two). This fact can be considered as an independent confirmation
of the better ``precision'' of the abundances calculated using WM technique. 

\begin{figure*}
\begin{center}
\begin{tabular}{c}
\includegraphics[angle=0,width=0.65\linewidth]{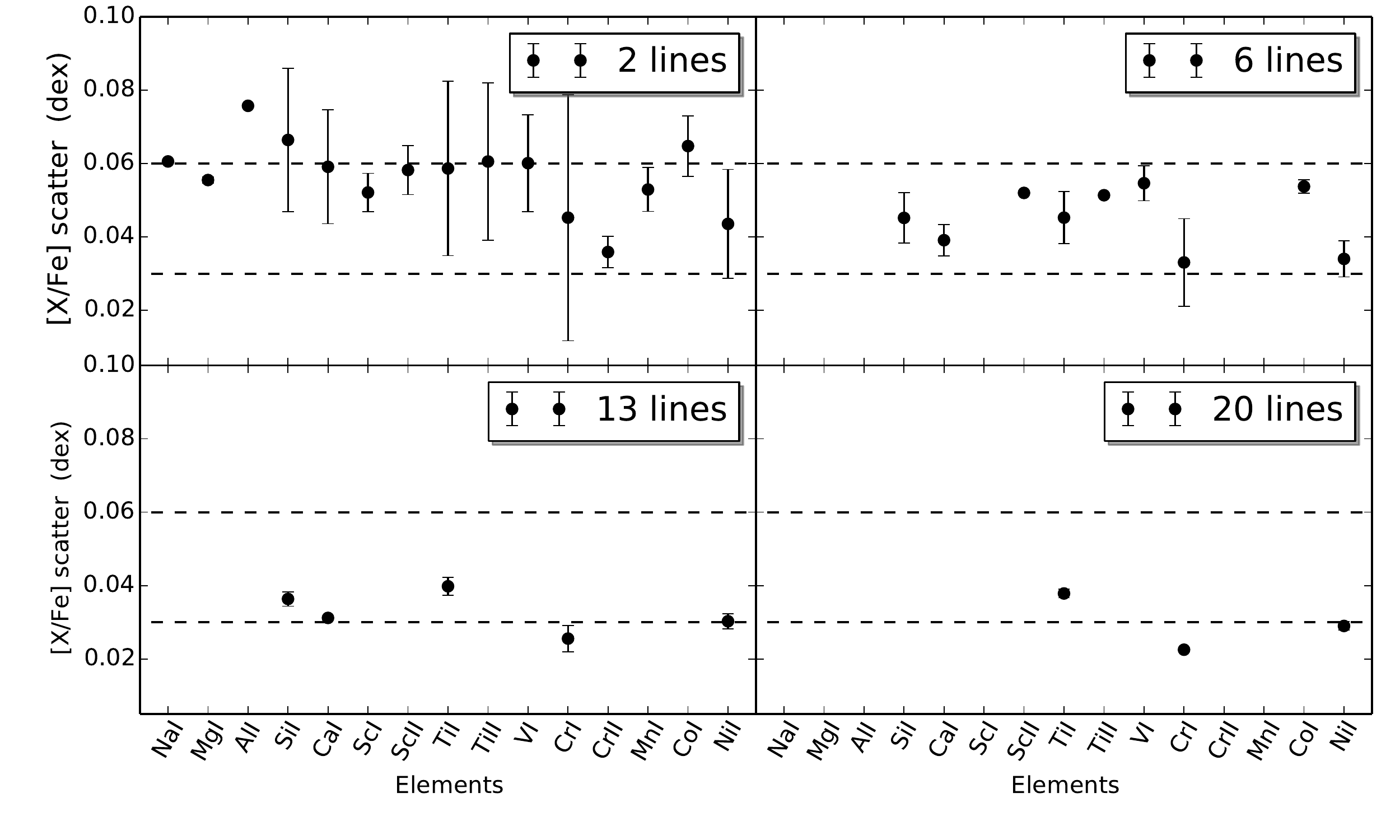}
\end{tabular}
\end{center}
\vspace{-0.5cm}
\caption{[X/Fe] star-to-star scatter for solar-analogs with solar-metallicity. The [X/Fe] scatter is derived by using 2, 6, 13 and 20 lines.
The error bars show the standard deviation of the scatter calculated using different combinations of lines.
The dashed lines, which represent [X/Fe] = 0.03 and 0.06 dex, are just to make the comparison of the [X/Fe] scatters between the elements visually easier.}
\label{fig_elements_elfe_scatter}
\end{figure*}

It is also very interesting to note that some individual lines can introduce a very large scatter while others provide very small one. The results of this test can be
used to rank the spectral lines according to the [X/Fe] scatter they provide. This can be used as a ``new'' method to eliminate outliers and select the best possible
lines. For example, there is one Ca line ($\lambda$5261.71) that clearly shows a very large scatter (0.16 dex) compared 
to the rest of 12 lines (on average $\approx$0.06 dex). It is interesting and important to note, that the average difference of the [Ca/H] abundance derived 
by using this line from the mean
abundance is very small, but again with a large dispersion $<$$\Delta$[Ca/H]$>$ = 0.05$\pm$0.15 dex, which means that the line does not show systematically higher or lower 
abundance when compared to that derived with the remaining lines. If all the 1111 stars are considered, then this difference becomes smaller, and negative 
$<$$\Delta$[Ca/H]$>$ = -0.02$\pm$0.15 dex. 
Our analysis of the [Ca/Fe] versus T$_{eff}$ for this line shows a very weak trend (0.05 dex/1000K) and 
particularly large scatter at low temperatures. However we found that the deviation of the Ca abundance of this line from the mean Ca abundance depends on the EW.
The average EW of this line is 104$\pm$23  and 116$\pm$45 m\AA{} for the solar analogs and for all the stars, respectively. 
When the EW is greater than 100 m\AA{}, the deviation increases significantly.

Similar to the discussed Ca line, we found some lines for the other elements that show distinguishably large dispersion. We provide the ranked list of 
all the lines ordered by the scatter size\footnote{The table is available at the CDS.}. 

Using the [X/Fe] scatter for all the elements and for different number of lines,   we compared the [X/Fe]
star-to-star scatter for different elements using the same number of lines in Fig.~\ref{fig_elements_elfe_scatter}.
We plot the scatter derived using 2, 6, 13, and 20 lines because these numbers are
those that better match the number of lines available: maximizing the number of lines and elements in the panels. 
For example, all the elements have at least two lines, and there is only one element that has a number of lines between 6 and 13. 

The top-left panel of Fig.~\ref{fig_elements_elfe_scatter} shows that when only two lines are used for all the elements to calculate [X/Fe], almost all the elements show a similar 
scatter of about 0.06 dex. Aluminum shows the largest, and Cr and Ni show the smallest scatters. However, one can also see that depending on the combinations of lines
the [X/Fe] scatter can be different for the same element (the error bar in the plot). The other three panels, that provide information which is more 
statistically significant since it is based on larger number of lines, show that from elements that have at least six lines, Ti, V, ScII, and Co show the largest scatter.
Again, Cr and Ni show the smallest scatter. We note that although the obtained differences in [X/Fe] scatter between elements are not large, they are based
on a large sample and thus can be considered statistically significant.

The decrease of the [X/Fe] scatter with the number of lines means that a fraction of the observed scatter does not have astrophysical origin. 
Table 3 of \citet[][]{Adibekyan-12} provides the average error of the [X/Fe] ratios for the same sample of stars. The table shows that the average error varies 
from 0.01 to 0.03 dex. For the elements that have at least 13 lines (SiI, CaI, TiI, CrI, and NiI) the average error on [X/Fe] is 0.01 dex.

Our results show the importance of the initial selection of the lines, especially when the number of lines is small. By carefully selecting lines
for individual stars with a given set of stellar parameters and a given quality of the spectra, one can derive precise chemical abundances even when
the number of lines is small and have small [X/Fe] scatter, as already demonstrated by e.g., \citet[][]{Ramirez-09} and \citet{Jonay-10} for solar analog stars. 
However, when dealing with large number of stars with different combinations of stellar parameters and quality of the spectra, it is not realistic to control abundances of each individual 
line in each individual star.

\section{Summary and conclusion}		\label{sec:conclusion}

In this paper, we used a large sample of FGK stars \citep[][]{Adibekyan-12} to study the dependence of precision of chemical abundances on the number of 
lines and how it affects the [X/Fe] star-to-star scatter at solar metallicities. We explored different techniques to calculate the mean abundance and minimize 
the effect of possible outliers when several spectral lines are available for an element.

From our tests we conclude and recommend to use the WM  (instead of any outlier removal technique) when several lines are available at hand. 
As a weight, the distance from the median abundance can be effectively used, as demonstrated.

Selecting only solar-analogs with metallicities similar to that of the Sun by 0.10 dex, we showed that [X/Fe] scatter strongly depends on the number of lines suggesting
that one should be cautious when comparing star-to-star abundance dispersion of elements which abundances were derived using different number of lines.
The decrease of scatter with the number of lines suggests that some fraction of the observed scatter has non-astrophysical nature. A large number
of lines is needed to reduce the precision induced scatter.

The comparison of the [X/Fe] scatter for different elements using the same number of lines show that most elements show a very similar dispersion.
The largest scatter among the elements studied in this work was found for Na, Al, Ti, V, ScII, and Co, while Cr and Ni show the smallest scatter. 
The similarity and differences in [X/Fe] scatter between the elements have different/similar nucleosynthesis production sites \citep[see e.g.][]{Nomoto-13}. 

Our group is currently working on the derivation of abundances of volatile (C and N) and r- and s-process elements 
(Su\'{a}rez-Andr\'{e}s et al, in prep; Delgado-Mena et al, in prep). When the data is ready a  similar analysis will be done for these elements to
select the elements that are the most informative for chemical tagging.

\begin{acknowledgements}
This work was supported by Funda\c{c}\~ao para a Ci\^encia e a Tecnologia (FCT) through the research grant UID/FIS/04434/2013. 
P.F., N.C.S., and S.G.S. also acknowledge the support from FCT through Investigador FCT contracts of reference IF/01037/2013, IF/00169/2012, 
and IF/00028/2014, respectively, and POPH/FSE (EC) by FEDER funding through the program ``Programa Operacional de Factores de Competitividade - COMPETE''. 
PF further acknowledges support from Funda\c{c}\~ao para a Ci\^encia e a Tecnologia (FCT) in the form of an exploratory project of reference IF/01037/2013CP1191/CT0001. 
V.A. and E.D.M acknowledge the support from the Funda\c{c}\~ao para a Ci\^encia e Tecnologia, FCT (Portugal) in the form of the grants 
SFRH/BPD/70574/2010 and SFRH/BPD/76606/2011, respectively. JPF acknowledges the support from FCT through the grant reference SFRH/BD/93848/2013. 
MO acknowledges support by Centro de Astrof\'isica da Universidade do Porto through grant with reference of CAUP-15/2014-BDP. 
JIGH acknowledges financial support from the Spanish Ministry of Economy and Competitiveness (MINECO) under the 2011 Severo Ochoa Program MINECO SEV-2011-0187. 
This work results within the collaboration of the COST Action TD 1308.

\end{acknowledgements}

\bibliography{adibekyan_bibliography}

\Online

\begin{appendix}

\section{Dependence of abundances on the number of lines: the case of Ni}  \label{app:a}

The data used in this work was taken from \citet[][]{Adibekyan-12}, which provides chemical abundances for 12 iron-peak and $\alpha$-capture elements (15 
ionized or neutral species).

The lines used in this work are based on the line-list of \citet{Neves-09}. From the VALD\footnote{Vienna Atomic Line Database} online database,
180 lines were carefully selected in the solar spectrum to be: not-blended,  have Equivalent Widths (EW) above 5 m\AA{} and below 200 m\AA{}, 
be located outside of the wings of 
very strong lines. Later on, the semi-empirical oscillator strengths for the lines were calculated by calibrating the $\log gf$ values to the solar reference of 
\citet{Anders-89}. Moreover, only ``stable'' lines which do not show high abundances dispersion (i.e., 1.5 times the $rms$) from the mean abundance for each element were selected. In this later test, 451 stars with 
wide range of stellar parameters and SNR were used. The selected 180 lines, were re-checked in \citet[][]{Adibekyan-12}, where several
lines were excluded because of the observed abundance trend [X/Fe] with the effective temperature. For more details about the selection of the lines 
we refer the reader to \citet{Neves-09} and \citet[][]{Adibekyan-12}. Our final line-list consists of 164 lines\footnote{This line-list was subsequently analyzed in
\citet{Adibekyan-15} to select a sub-list of lines suitable for abundance derivation for cool, evolved stars}.

\begin{figure*}
\begin{center}
\begin{tabular}{cc}
\includegraphics[angle=0,width=0.45\linewidth]{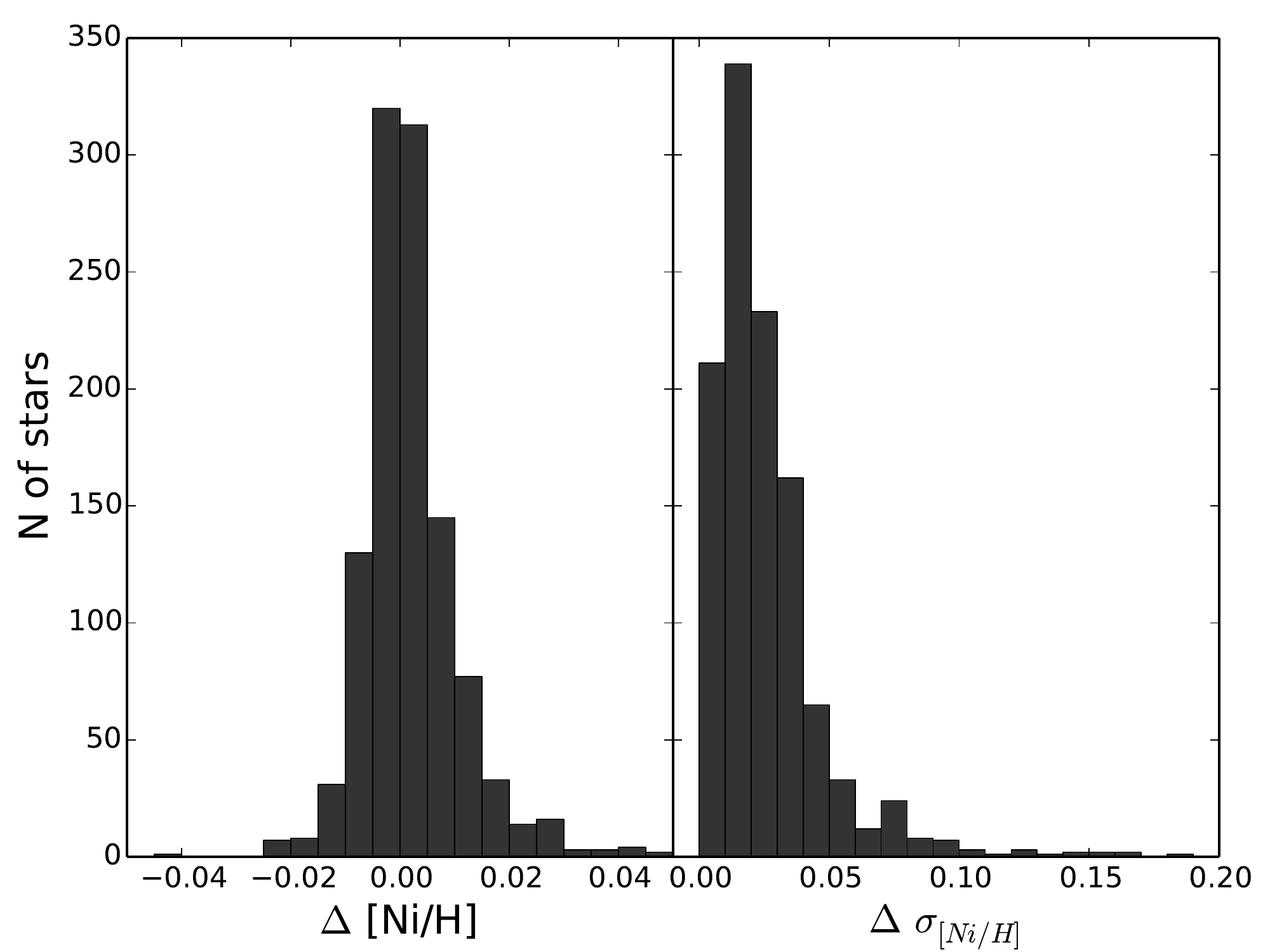}
\includegraphics[angle=0,width=0.45\linewidth]{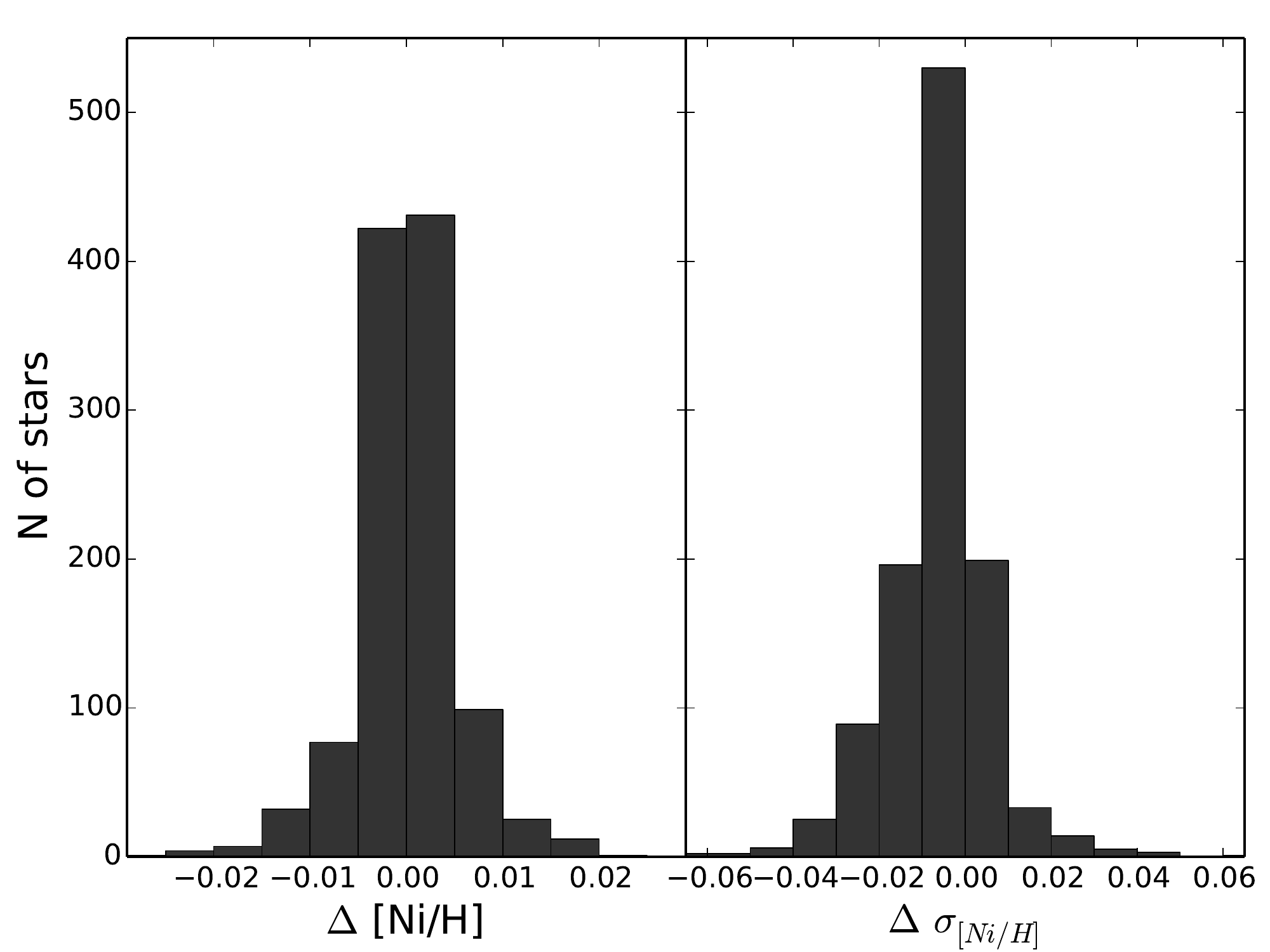}
\end{tabular}
\end{center}
\vspace{-0.5cm}
\caption{Difference in [Ni/H] and $\sigma_{[Ni/H]}$ when WM and AM without outlier removal methods are applied (left). The same as in the
left panel, but the parameters are derived using WM and MAD$_{e}^{iter}$ (with the threshold of median$\pm$3MAD) techniques.}
\label{fig_delta_ni_abund}
\end{figure*}

We stress that the main goal in this work is not to re-check the quality of the lines, nor to provide a range of parameter space (stellar parameters and SNR)
where each individual line can be safely and reliably used. Since different authors use different set of spectral lines and different atomic data for the lines,
for us it is more straightforward and scientifically interesting to discuss methods that can, in principle, effectively work for different line-lists and 
when applied on large datasets, as it is offen the case.

\begin{figure*}
\begin{center}
\begin{tabular}{cc}
\includegraphics[angle=0,width=0.45\linewidth]{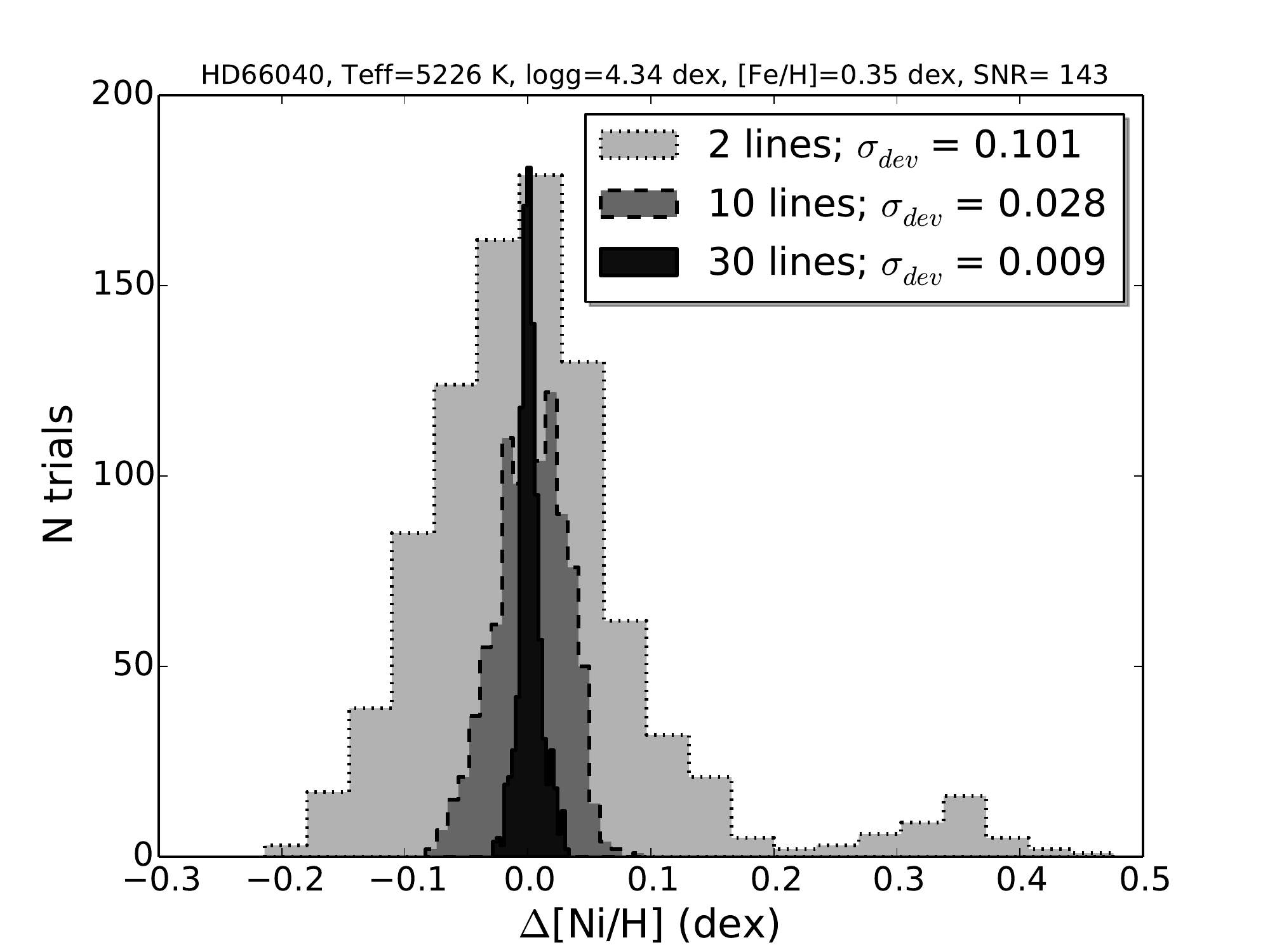}
\includegraphics[angle=0,width=0.45\linewidth]{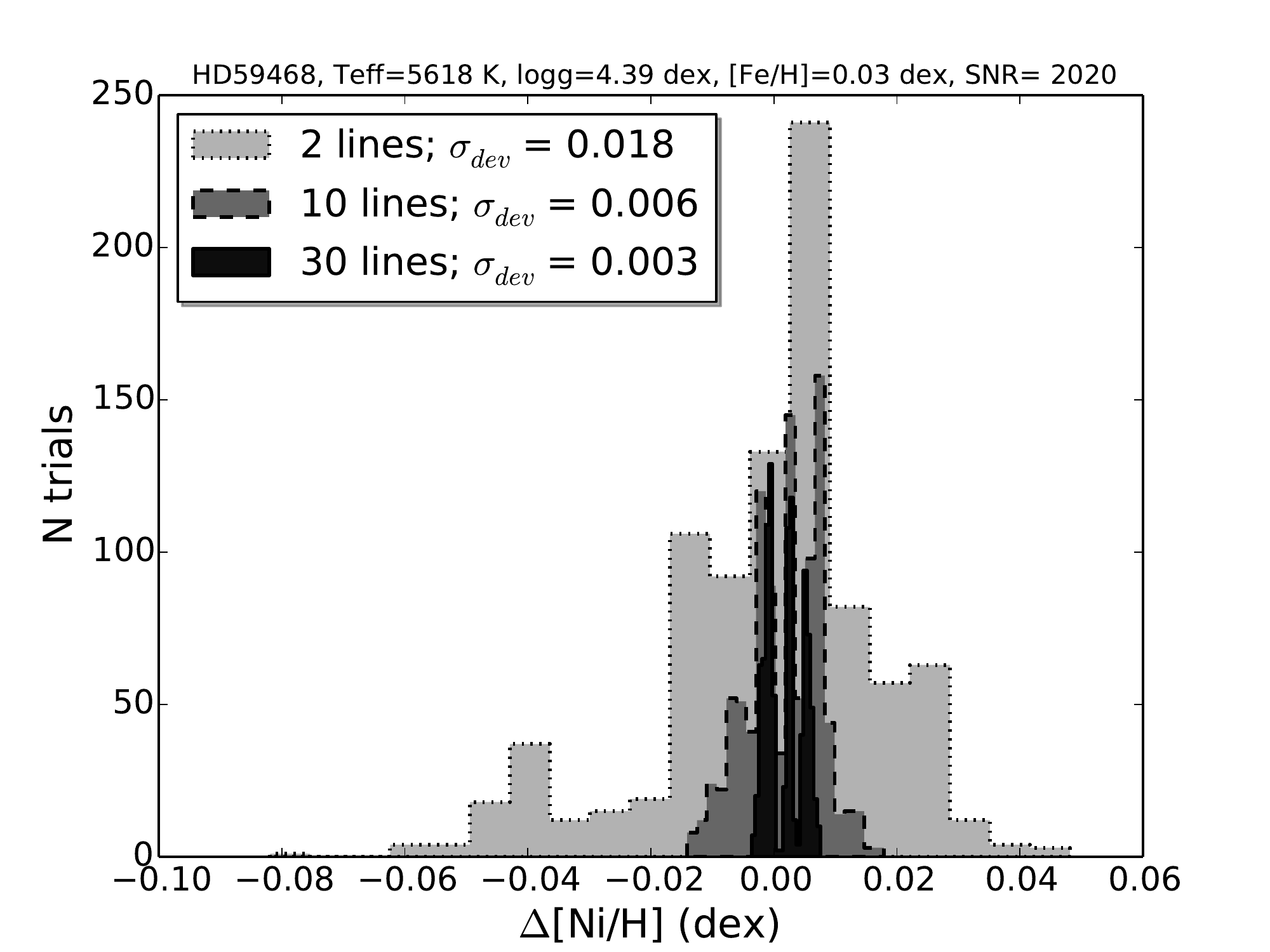}
\end{tabular}
\end{center}
\vspace{-0.5cm}
\caption{The difference between original Ni abundance and Ni abundances derived with only 2, 10, and 30 Ni lines.}
\label{fig_ni_abund_example}
\end{figure*}

\subsection{Comparing methods} 

In \citet[][]{Adibekyan-12} the final abundance for each star and element was calculated as the arithmetic mean (AM) of the abundances given by all lines detected
in a given star and element after a 2-$sigma$-clipping was applied. 
This is a standard, and widely used technique that allows to avoid the errors caused by bad pixels, bad measurements, cosmic rays, and other unknown 
localized effects. 
However, this type of ``outlier'' removal technique
depends on the threshold (2-$\sigma$ in our case) that is applied for which there is no clear prescription, or theoretical ground, and the choice ends up being very subjective.
The choice of threshold should also depend on the sample size.
A simple demonstration of this sample size dependence is presented by \citet[][]{Shiffler-88}, who showed that the possible maximum Z-score 
(number of SD a 
data-point is far from the mean) depends  (only) on the sample size and it is computed as (n-1)/$\sqrt[]{n}$. From this formula we get, 
 that the maximum deviation one can obtain in a 
sample of 5 points (lines) is 1.79-$\sigma$, i.e., no outliers can be identified in the data if 2-$\sigma$-clipping is applied. 
One can alternatively use median and median absolute deviation (MAD), which is expected to be less sensitive to outliers, or apply other outlier removal 
methods \citep[e.g.][]{Hodge-04, Iglewicz-93}. However, it is very difficult to choose a single method and a criterion that will efficiently work for 
samples of different size. Moreover, when a certain criterion is applied to remove possible outliers, some valid lines from the real distribution can be
removed as well. Finally, one should also bear in mind that most of the outlier removal methods are model-dependent assuming some distributions
for the real and outlier data.

\begin{table}
\centering
\caption{The difference in Ni abundances when WM method and other methods are applied for the derivation of Ni.}
\label{table-tests}
\begin{tabular}{ccc}
\hline 
Methods & Threshold & $\Delta$Ni (dex) \tabularnewline 
\hline 
\hline \vspace{0.1cm}
WM -- AM & -- & -0.0020$\pm$0.0096\tabularnewline

\multirow{3}{*}{WM -- Median-rule} & median$\pm$2IQR & 0.0001$\pm$0.0052 \tabularnewline
 & median$\pm$2.5IQR & -0.0002$\pm$0.0057 \tabularnewline \vspace{0.1cm}
 & median$\pm$3IQR & -0.0006$\pm$0.0062 \tabularnewline  

\multirow{3}{*}{WM -- $\sigma$-clipping} & median$\pm$2SD & 0.0003$\pm$0.0049 \tabularnewline
 & median$\pm$2.5SD & -0.0002$\pm$0.0054 \tabularnewline \vspace{0.1cm}
 & median$\pm$3SD & -0.0004$\pm$0.0062 \tabularnewline  
 
\multirow{3}{*}{WM -- MAD$_{e}$} & median$\pm$2.5MAD$_{e}$ & 0.0002$\pm$0.0049 \tabularnewline
 & median$\pm$3MAD$_{e}$ & 4.5$\times10^{-5}\pm$0.0057 \tabularnewline \vspace{0.1cm}
 & median$\pm$3.5MAD$_{e}$ & -0.0006$\pm$0.0062 \tabularnewline 

\multirow{3}{*}{WM -- MAD$_{e}^{iter}$} & median$\pm$2.5MAD$_{e}$ & 0.0004$\pm$0.0055 \tabularnewline
 & median$\pm$3MAD$_{e}$ & 0.0001$\pm$0.0052 \tabularnewline \vspace{0.1cm}
 & median$\pm$3.5MAD$_{e}$ & -0.0002$\pm$0.0057 \tabularnewline
  
\hline 
\end{tabular}
\end{table}

\begin{figure*}
\begin{center}
\begin{tabular}{cc}
\includegraphics[angle=0,width=0.45\linewidth]{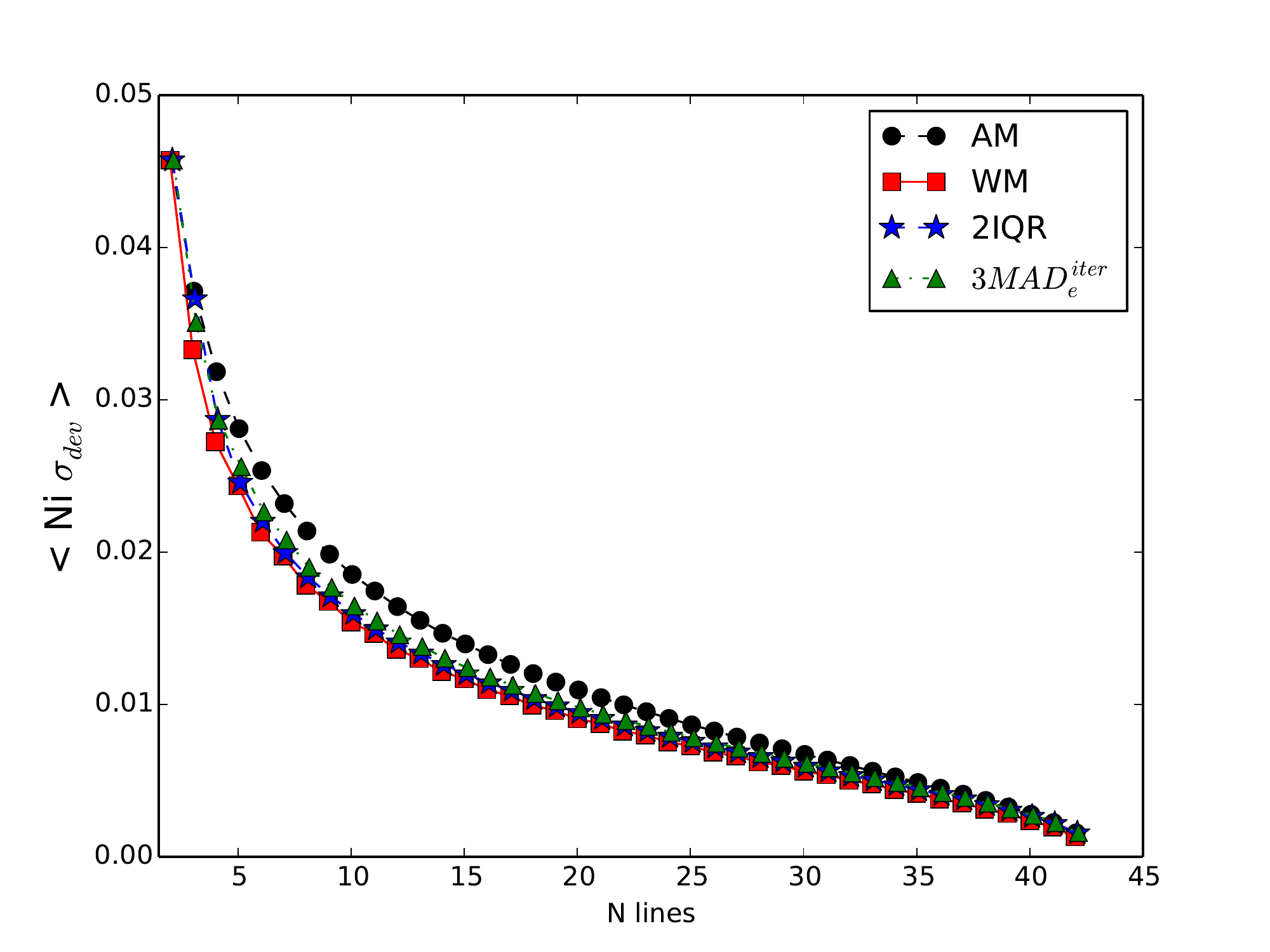}
\includegraphics[angle=0,width=0.45\linewidth]{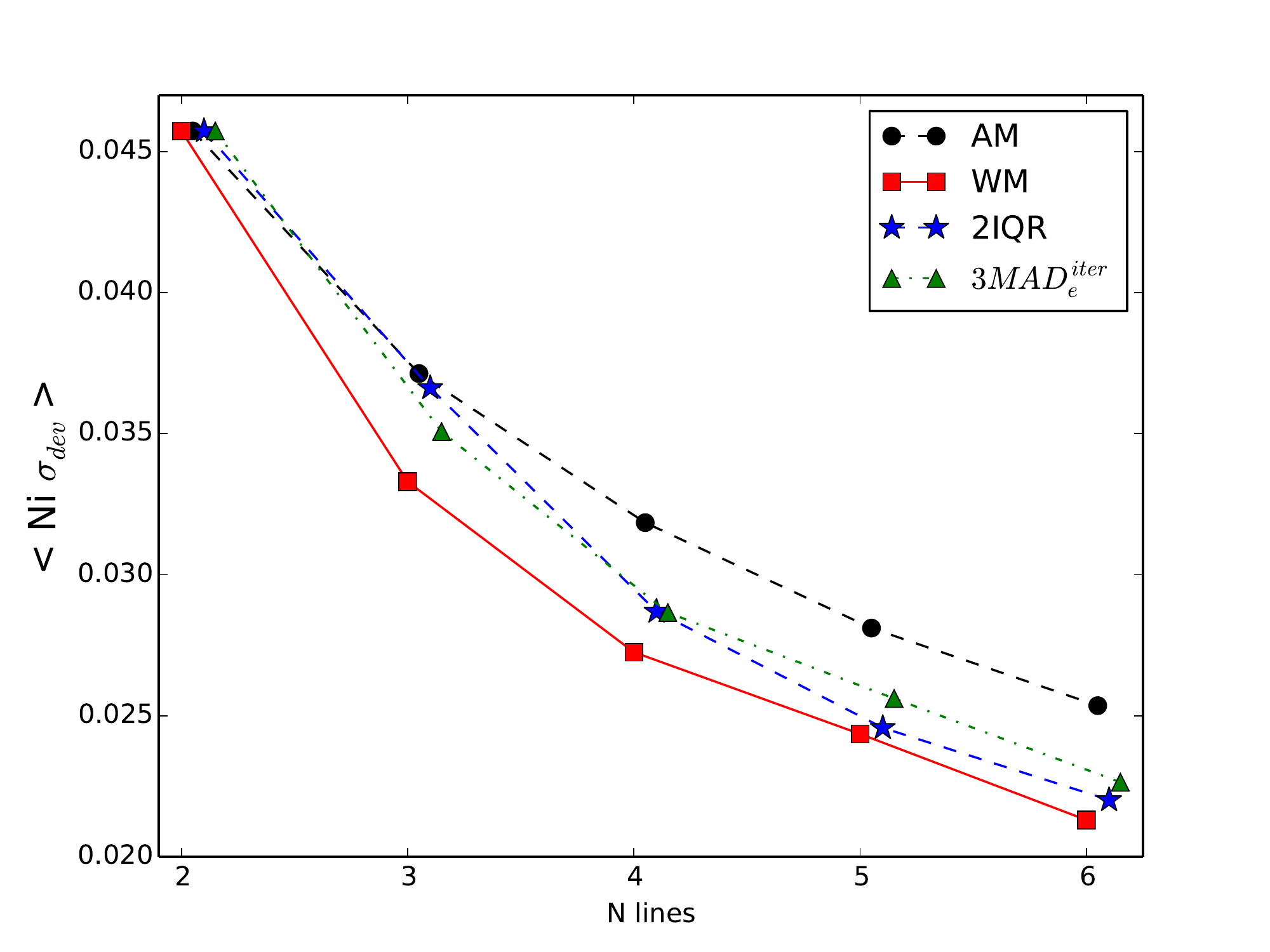}
\end{tabular}
\end{center}
\vspace{-0.5cm}
\caption{Average deviation from the original Ni abundance for 1111 stars versus number of lines that were used for the abundance derivations. The $right$
panel is the zoom of the left plot, only limited to six lines. Different techniques
that were used in the calculations are mentioned in the plots.}
\label{fig_delta_ni_nlines}
\end{figure*}

\begin{figure*}
\begin{center}
\begin{tabular}{c}
\includegraphics[angle=0,width=0.7\linewidth]{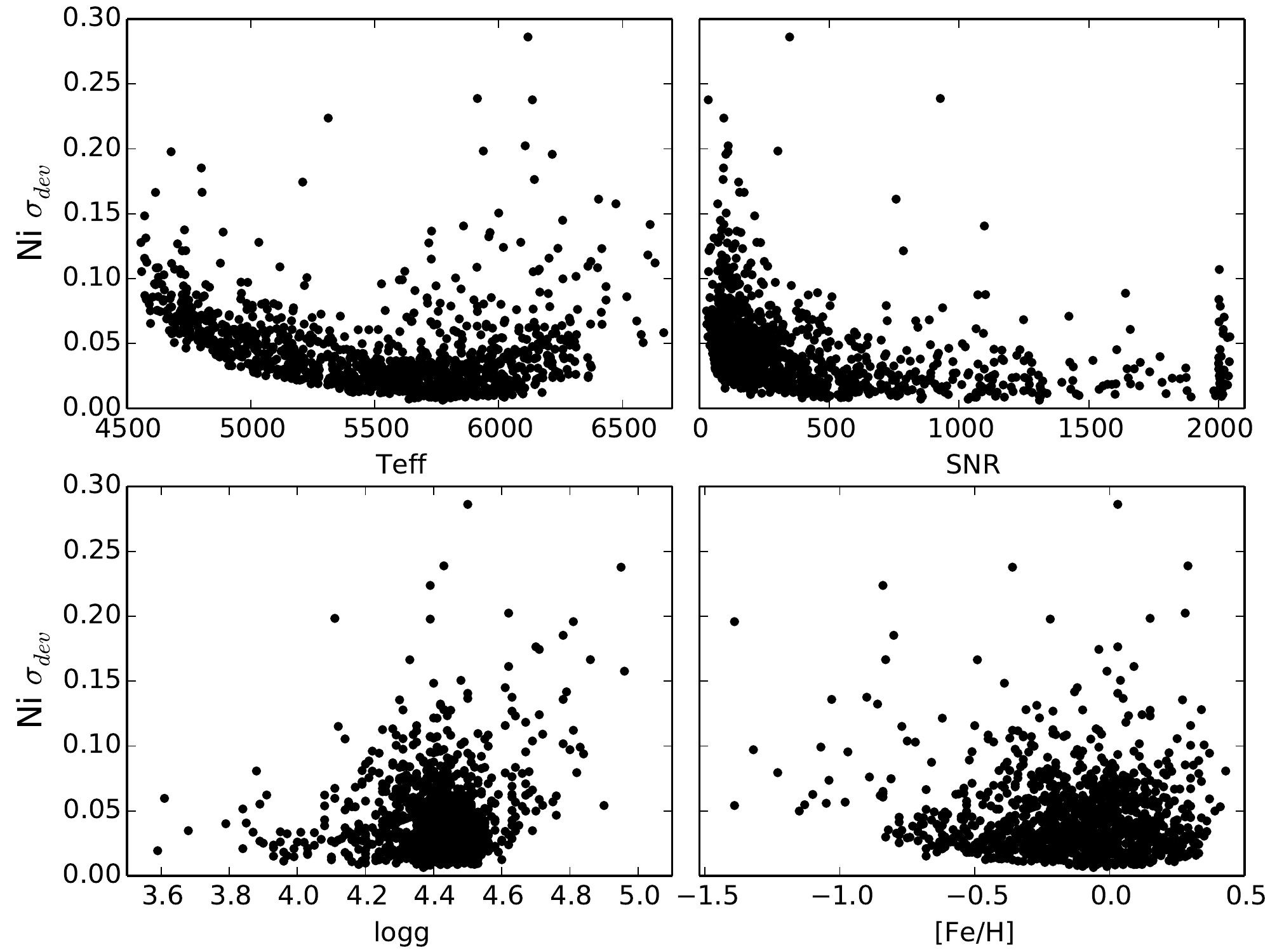}
\end{tabular}
\end{center}
\vspace{-0.5cm}
\caption{Dependence of the average deviation from the original Ni abundance for 1111 stars versus stellar parameters and SNR. The abundance deviation
represents the difference between original Ni abundance and Ni abundance derived only with 2 lines.}
\label{fig_ni_scatter_parameters}
\end{figure*}

To explore and choose the method that allows to derive the most precise final abundances of the elements, we 
selected Ni for our analysis because it has  the largest linelist (43 lines). By plotting the individual Ni abundances 
in the full sample of 1111 stars we noticed that 
many of the stars have Ni lines which show deviation from the average value by more than 3-$\sigma$. 
To understand if these lines are outliers or just extremes of the distribution  (a normal distribution is assumed here) we performed some simple calculations. 
If one assumes a normal distribution, then a 3-$\sigma$ corresponds to P = 0.003 probability. 
Since on average we derive Ni abundance from 43 lines, then the probability that we will have at least one ``outlier'' is of 43$\times$0.003=0.129. 
This means that among the 1111 stars we expect to have about 0.129$\times$1111=143 stars with one ``outlier'' line. 
However, the number of stars which have at least one ``outlier'' is 626. To estimate the probability of 
having that many stars with at least one ``outlier'' we used binomial probability distribution. 
The probability that more than 200 stars (any number above 200) can have an ``outlier'' is already 8$\times$10$^{-7}$. This means that, under our assumption
of Gaussian distribution of the abundances derived from different lines, some of the lines which show large dispersion ($>$ 3-$\sigma$) can be real outliers 
of different origin and are not just coming from the wings of the Gaussian distribution. We note, that the results of this test do not depend on the 
applied threshold (3-$\sigma$ in this case).

Fortunately, if the linelist is large, the possible outliers do not affect much the final (mean) abundance. We first tested  three different outlier removal methods,
namely $\sigma$-clipping \citep[e.g.][]{Shiffler-88}, modified Z-score \citep[][]{Iglewicz-93}, and median-rule \citep[][]{Carling-98} on our data. The modified Z-score method is similar to   
n$\times$ $\sigma$-clipping, but instead of mean and SD,  the median and MAD$_{e}$\footnote{MAD$_{e}$ = 1.483$\times$MAD, and is equal to SD for large normal data}
are used. Median-rule is a modification of Tukey's (boxplot) method \citep{Tukey-77} and defines outliers as points that lies further than median$\pm k\times$IQR,
where IQR is the interquartile range. We varied the $k$ to  $\{2,2.5,3\}$ for the $\sigma$-clipping and median-rule, and 
$k$ = $\{2.5,3,3.5\}$ for the modified Z-score. The selected values of $k$ are within the intervals suggested in the above cited references. 

A potential difficulty that one faces when trying to remove outliers is the so-called masking and swamping effects - removal of one outlier changes the ``status'' of the other data
points \citep[][]{Acuna-04}. This means that it is advisable to remove one outlier at the time and apply the criteria recursively.
However, it is not obvious when the outlier removal criteria should be stopped (the problem exists also when the outliers are removed at once).
Two approaches were considered in our tests when modified Z-score method was used: i) remove all the outliers at once (we call it MAD$_{e}$ technique in the remainder of the paper), 
and ii) remove one outlier at a time and then apply the criterion again iteratively (hereafter we call it MAD$_{e}^{iter}$ technique). 
For the second approach we allowed maximum number of 10 iterations, although in most of the cases, a lower number of  
iterations were needed (depending, of course, on the threshold accepted).

Outlier removal is not the only method used to characterize an underlying distribution in a dataset. An example is the  
weighted least-squares regression to minimize the effects of outlier data \citep[][]{Rousseeuw-87}. 

The last method that we use to calculate the
final abundance and its line-to-line scatter is the WM and weighted SD. As a weight we used the  (inverse) distance from the median value in
terms of SD and then binned it. Using MAD and SD in the calculations of the weight on the average give very similar results, but if the values 
of more than the half of the points 
(lines) are the same (this can happen when the number of lines is small), then MAD is by definition zero, and cannot be used to calculate the weight. 
Since the distance of the median point from the median is zero, the weight of that line
would be infinite. To avoid giving a very high weight to the points that are initially close to the median (the final value would by construction be very close 
to the median), we decided to bin the distances with an interval of 0.5SD. 
E.g., a 0.5$\times$SD weight was given to the lines that are at the distance from  0 to 0.5 SD from the median. Similarly, a 1$\times$SD weight was given 
to the lines lying at the distances of 0.5 to 1$\times$SD, and so on.

The results of our tests are summarized in the Table~\ref{table-tests}. The test showed that all the outlier removal methods give a mean, 
final abundance similar to the one of the WM. Since the number of lines is relatively large, the impact of possible outliers is small, 
and all the values were also similar to the abundance calculated by the AM of all the points.
However, we note that when the lowest thresholds were set to remove outliers, some stars due to the large number of removed ``outliers'', showed deviations
in the final abundance from the mean abundance derived from different methods. Another important point to stress is that 
when  outlier removal methods were applied with low thresholds, the line-to-line scatter (which is usually used as an error estimate of the 
final abundance) was usually small, as expected.

From these tests (and further tests presented next in this work), we concluded that the best way to calculate the final abundance and its error is to use the 
WM. In this case, the weight of real outliers (extremes) is small, and the final abundance is not affected. With this approach, we also do not reduce 
the scatter by artificially removing points from the distribution. In Fig.~\ref{fig_delta_ni_abund},  we plot the distribution of the Ni abundance
and its error (line-to-line dispersion) differences  when WM and AM method is applied (left plot), and when WM
and MAD$_{e}^{iter}$ (median$\pm$3MAD) criteria is applied (right plot). From the plot and table it is very clear that when the number of 
lines is large, different outlier removal (or not)  methods provide very similar results for Ni abundances, 
however the error associated to these values depends on the method. In particular,  Fig.~\ref{fig_delta_ni_abund} shows (left plot) that line-to-line scatter of [Ni/H]
 is always larger when the Ni abundance is calculated by the AM than when the WM method is used  (the difference in $\sigma_{[Ni/H]}$ is always positive).
 The right panel of the same figure, shows that the difference in $\sigma_{[Ni/H]}$ when Ni abundance is calculated by WM and MAD$_{e}^{iter}$, is usually small
 and can be both positive and negative.

A word of caution should be added at this point. In the methods that we tested to remove ``outliers'' and in the WM technique we assume that the distribution of 
the abundances (or the distribution of the errors on abundances) is symmetric\footnote{Note that for the WM method there is no assumption on the normality of the
distribution of the errors of abundances, while some outlier removal methods based on this assumption.}. However, as it was shown in \citet{deLis-15} for very weak lines with an assumption of
LTE (local thermodynamic equilibrium) the distribution of uncertainties of abundances is asymmetric. The authors also showed that this effect depends on the SNR and 
is negligible for lines with EW greater than 8m\AA{} regardless of SNR. However, since in all the methods are based on the same hypothesis, the WM technique remains
favorable for us.

\subsection{Abundance precision dependence on the number of lines}

To evaluate the impact of the number of lines (that one uses for abundance derivations) on the abundances, we did the following simple tests. For each star in the sample,
we randomly drew $N$ Ni lines ($N$ = 2,3,...,42) and calculated the Ni abundance. We used the above mentioned WM technique for the calculation of the
abundances. Then we compared the resulting abundances with the supposed Ni abundance value 
(derived by using all the 43 lines available and the WM technique). If the number of possible combinations is less than 1000,
we considered at all the possible combinations of lines, otherwise we drew $N$=1000 random, but different combinations of lines\footnote{We note that 1000 is a sufficiently high
number of combinations and our tests showed that increasing this number by a factor of 100 has negligible impact on the results.}.

In Fig.~\ref{fig_ni_abund_example}, we plot an example (for two stars) of the distribution of the differences in Ni abundances ($\Delta$[Ni/H]) when three different number of 
Ni lines (2, 10, and 30 lines) and all the available lines are used. The stars have different stellar parameters and different SNR in the spectra. 
The plot shows that when the number of lines is increased the abundance difference gets smaller. It also shows that
while most of the cases/trials the $\Delta$[Ni/H] is close to zero, it is possible to obtain very large differences when only two lines are used (even for very high SNR data).

We did the aforementioned computations for all the 1111 stars and for each number of lines we calculated the standard deviation 
of $\Delta$[Ni/H] distribution - $\sigma_{dev}$.

In Fig.~\ref{fig_delta_ni_nlines}, we plot the dependence of the average of the $\sigma_{dev}$ for all 1111 stars as a function of the number of lines.
In the plot, we  only limited ourselves with  examples of four techniques with different thresholds in order not to overload the figure, 
while applying all the techniques and thresholds presented in Table~\ref{table-tests}. Moreover, since in these tests the size of the sample (lines)
varies, we decided to test also lower outlier removal thresholds: $k$ = 1.5 for $\sigma$-clipping and median-rule methods, and $k$=2 for modified Z-score methods).

Fig.~\ref{fig_delta_ni_nlines} shows the range of possible deviations (1$\sigma$ 
deviation if the distribution was a Gaussian) from the original value for a given random star when a randomly draw $N$ lines are used.
It  clearly shows that the deviation decreases very steeply with the number of lines and becomes 
smaller than 0.01 dex when more than 15 lines is used.

On the right panel of Fig.~\ref{fig_delta_ni_nlines}, we show that, for a number of lines less than or equal to six, there is a subtle difference
between different outlier removal techniques. It clearly shows that the smallest average  deviation is obtained when the WM is used. 
We note, that other tests with different thresholds show similar results. The low thresholds for outlier removal techniques give results closer to 
that obtained by using WM for small number of lines. However, when low thresholds are considered for a large number of lines, due to high number of excluded
lines, the final results deviate from the abundances obtained by using WM method.
For the remainder of the paper we use abundances calculated by the WM method, if another method is not specified.
Here we should stress again that we plot the possible deviations of Ni abundances averaged for 1111 stars. 
While these average values are small, the deviations for individual
stars can be very significant (as demonstrated in Fig.~\ref{fig_ni_abund_example}). 

It is natural to expect that the observed deviations should depend chiefly on the quality of the data (e.g. SNR) and also on the atmospheric parameters of the stars.
This is because e.g. spectral lines in cooler stars spectra are usually more blended, and also because e.g. different lines form at different layers of the 
atmospheres and have different sensitivities to the non-LTE effects.
In Fig.~\ref{fig_ni_scatter_parameters}, we plot for the case 
in which only two lines were used the dependence of the average $\sigma_{dev}$ on the stellar atmospheric parameters and on the SNR. 
The plot shows that there is only a strong and clear dependence on T$_{eff}$. This result is expected since at low temperatures the spectra of cool stars
are crowded and line blending plays a stronger role.  
Lowest metallicity stars and stars with the lowest SNR also show somewhat larger deviations. 
It is interesting to note that even if the SNR is very high, depending on
stellar parameters, it is possible to obtain a Ni abundance up to 0.1 dex different from the original abundance when only two Ni lines are used.

\section{[X/Fe] star-to-star scatter: dependence on the number of lines}

\begin{figure*}
\begin{center}
\begin{tabular}{c}
\includegraphics[angle=0,width=0.7\linewidth]{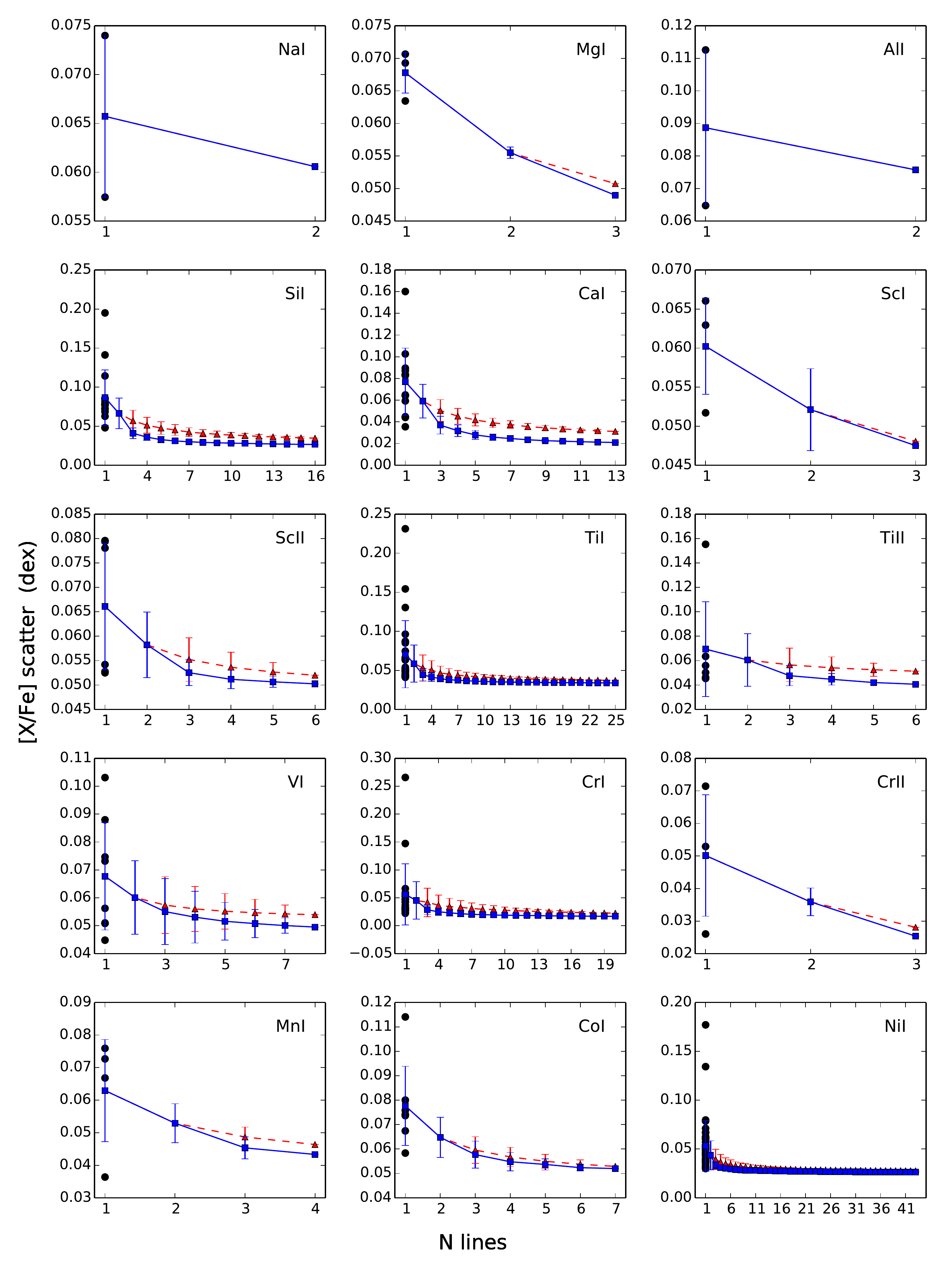}
\end{tabular}
\end{center}
\vspace{-0.5cm}
\caption{Dependence of [X/Fe] star-to-star scatter for solar-analogs with [Fe/H] = 0.0$\pm$0.10 dex on the number of lines. Red triangles show the scatter when
the individual abundances are calculated as an AM and the blue squares indicate the scatter in [X/Fe] when the WM method was used for 
the abundance derivation. The black dots show the [X/Fe] scatter for each individual line that was used to derive [X/H]. The error bars indicate the dispersion
of possible combinations of the lines.}
\label{fig_elements_elfe_scatter_nlines}
\end{figure*}

\end{appendix}

\end{document}